\documentclass[aps,amsmath,amssymb,reprint,citeautoscript,groupedaddress,shortbibliography,superscriptaddress]{revtex4-2}

\usepackage{amssymb,amsmath,mathtools,graphicx}
\usepackage{xspace}  % to get correct spacing after \newcommand substitutions, http://tex.stackexchange.com/a/17731
\usepackage{siunitx} % for proper units
\usepackage{xcolor}

\newcommand{\dif}{\mathrm{d}}
\newcommand{\vek}[1]{\boldsymbol{#1}} 

\begin{document}

\title{Leaf water storage and robustness to intermittent drought: A spatially explicit capacitive model for leaf hydraulics}
\author{Yongtian Luo}
\email{yongtian@sas.upenn.edu}
\affiliation{Department of Physics and Astronomy, University of Pennsylvania}
\author{Che-Ling Ho}
\affiliation{Department of Biology, University of Pennsylvania}
\author{Brent R. Helliker}
\affiliation{Department of Biology, University of Pennsylvania}
\author{Eleni Katifori}
\email{katifori@sas.upenn.edu}
\affiliation{Department of Physics and Astronomy, University of Pennsylvania}

\begin{abstract}
Leaf hydraulic networks play an important role not only in fluid transport but also in maintaining whole-plant water status through transient environmental changes in soil-based water supply or air humidity. Both water potential and hydraulic resistance vary spatially throughout the leaf transport network, consisting of xylem, stomata and water-storage cells, and portions of the leaf areas far from the leaf base can be disproportionately disadvantaged under water stress. Besides the suppression of transpiration  and reduction of water loss caused by stomatal closure, the leaf capacitance of water storage, which can also vary locally, is thought to be crucial for the maintenance of leaf water status. In order to study the fluid dynamics in these networks, we develop a spatially explicit, capacitive model which is able to capture the local spatiotemporal changes of water potential and flow rate. In electrical-circuit analogues described by Ohm's law, we implement linear capacitors imitating water storage, and we present both analytical calculations of a uniform one-dimensional model and numerical simulation methods for general spatially explicit network models, and their relation to conventional lumped-element models. Calculation and simulation results are shown for the uniform model, which mimics key properties of a monocotyledonous grass leaf. We illustrate water status of a well-watered leaf, and the lowering of water potential and transpiration rate caused by excised water source or reduced air humidity. We show that the time scales of these changes under water stress are hugely affected by leaf capacitance and resistances to capacitors, in addition to stomatal resistance. Through this modeling of a grass leaf, we confirm the presence of uneven water distribution over leaf area, and also discuss the importance of considering the spatial variation of leaf hydraulic traits in plant biology.
\end{abstract}
\maketitle

\section{Introduction}
\label{sec:intro}

Fluid flows in the plant vascular tissue system, which consists of xylem vessels for water transport and phloem vessels for the transport of photosynthetic products from leaves, are by no means isolated from other plant tissues. This is especially prominent in leaf hydraulic networks, which are typically the terminal portions of water flow through xylem. The xylem vessels making up these networks not only connect to phloem through leaf tissue, but also deliver water to an extra-xylery network of living cells from which water is evaporated and  transpired to the atmosphere through leaf-surface pores (stomata)\cite{Sack2006,Stroock2014}. Photosynthetic carbon assimilation requires stomata to remain open for the exchange of carbon dioxide with air, while transpiration simultaneously leads to a large sum of water loss, resulting in  water-use efficiency ($\text{CO}_2$ uptake per water molecule loss) as low as $1/500$ \cite{Taiz2002}. Implementation of water stress by a shortage of water source at the leaf base or decreasing atmospheric humidity around the leaf will cause stomata to close, thus suppressing transpirational water loss, but also reducing photosynthesis \cite{Brodribb2009,Choat2018}. The maintenance of leaf water status, which changes spatially in the xylem hydraulic vascular network, is therefore critical to keeping stomata open and sustaining photosynthesis. 

It has been proposed that water storage functions of certain cells in leaf tissue could help maintain leaf water status, and support the resilience and survival of a plant experiencing water stress \cite{Tyree1991,Jones2013}. In a leaf of a succulent plant, for example, water-storage parenchyma cells play the role of hydraulic capacitors, storing water when water supply is sufficient and providing water to sustain water status under stress \cite{Smith1987}. In a grass leaf, bulliform cells, water-storage parenchyma, as well as vascular bundle sheaths, could play a similar role \cite{raven2005biology}.  Previous work on the water-storage capacitance performed by these cells was from the perspective of whole leaf or plant body, such as in a lumped-element model using electrical-circuit analogues, where a whole system-wide capacitor is used as well as other whole-system elements including resistors \cite{Jones1978}. These reservoir cells, however, are distributed along water pathways in the network, which means capacitance is spatially dependent and could affect the water status locally. Transpiration processes, on the other hand, would also occur locally through stomata all over the leaf surface, making vessels in the network behave as leaking pipes. The competing effects of transpiration and water storage under stress will thus be more appropriately investigated in terms of spatially explicit network systems. In grass leaves, the unbalanced distribution of water content from leaf base to tip is illustrated by the fact that the area near tip is disproportionately disadvantaged and dries out faster than the area near base (water source) when subject to water shortage or even a transient change in atmospheric humidity. 

The plant or leaf water status, generally described by water potential $\psi$ which is regarded as the driving force of water flow, has been theoretically studied by two classes of models. In both classes, the water transport through leaf xylem is treated as laminar following the Hagen-Poiseuille law, in which the hydraulic resistance of a xylem vessel is equal to the water potential difference between its two ends divided by the flow rate \cite{Tyree1991,Altus1985}. The first type of models considers the small-scale spatial variations of leaf vascular networks, by implementing a network system consisting of only resistors, while ignoring the water-storage capacitance \cite{Cochard2004,Katifori2018}. In the second modeling type, large-scale tissue or organ-level properties including both resistance and capacitance are considered, while ignoring any spatially explicit architecture within a leaf. In such a model, the total transpiration rate going through the leaf is commonly controlled by a current source and can be adjusted to an arbitrary constant value, without any physical input or mechanistic explanation. Here, we bridge these two classes of models by developing a spatially explicit leaf hydraulic network model with local capacitance. While our model is general and can be applied to any type of vascularized leaf, we focus on grass leaf examples as they almost ubiquitously have water storage cells (bulliform cells), and the parallel vein structure of grasses leads to water being lost throughout the length of the blade. While this too occurs in dicots over short distances \cite{Zwieniecki2002}, the process occurs along the entirety of a grass leaf. By conducting computation and simulation on a uniform grass leaf model, we study the dependence of transpiration rate on leaf hydraulic traits and water potentials in the environment. We illustrate and discuss how capacitance increases the robustness of leaf in a changing environment and maintains leaf water status, so that stomata can remain open, potentially sustaining photosynthesis along the entire leaf blade.

\section{The spatially explicit model of capacitive leaf hydraulics}

\begin{figure*}[hbt!]
\centering
\includegraphics[width=0.8\textwidth]{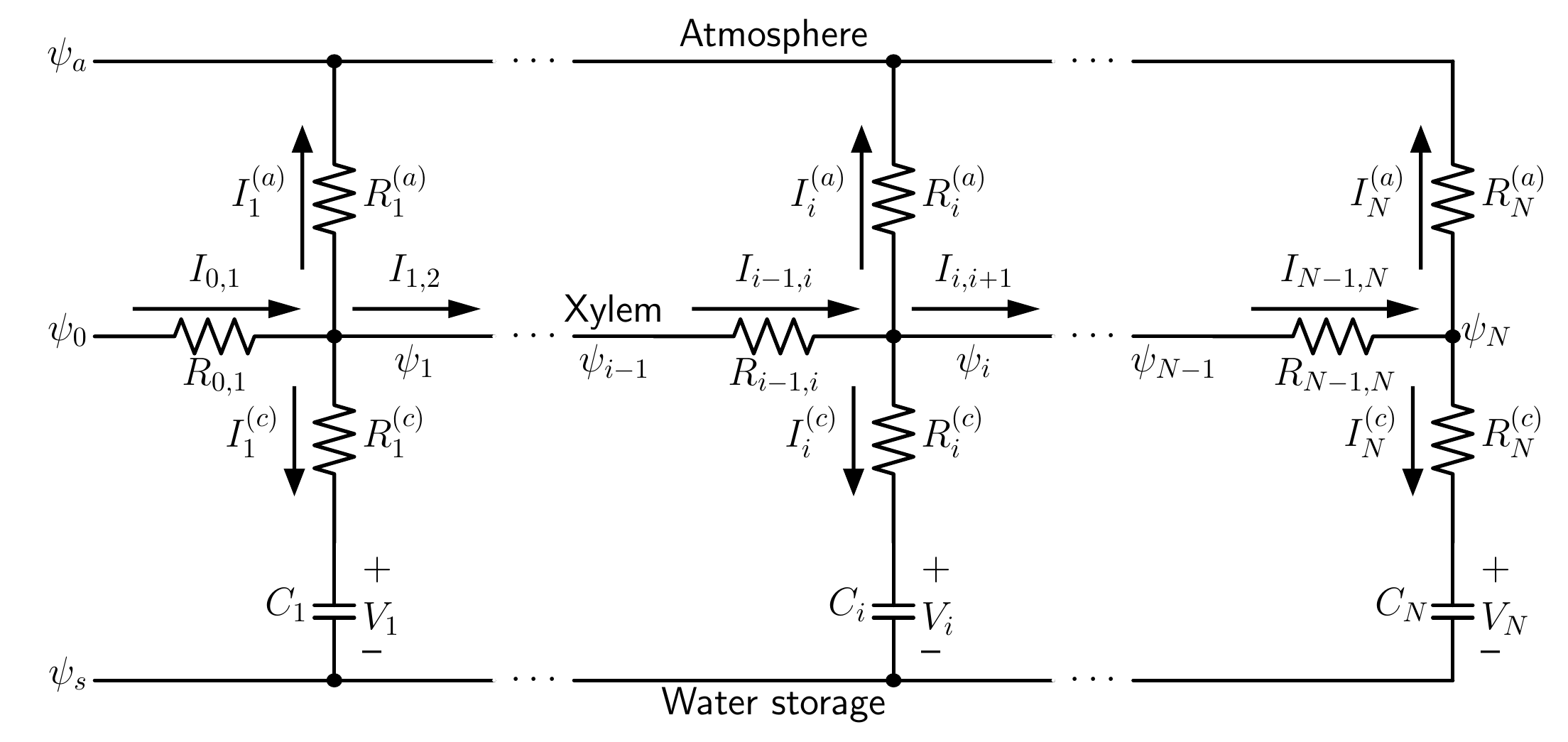}
\caption{A one-dimensional capacitive network model, where $\psi_a$ is the water potential in the atmosphere, $\psi_s$ is the baseline potential of water storage, $\psi_0$ is the water potential at the base, and $\psi_i$ ($i=1,2,\dots,N$) are water potentials at different nodes of the xylem. Currents $I_i^{(a)}$ are transpiration currents through stomata, and $I_i^{(c)}$ charge capacitors $C_i$, which represent leaf water storage function.}
\label{uniform}
\end{figure*}

We use a capacitive electrical circuit analogue to model the spatial variation of hydraulic traits of a simple plant leaf model (such as a monocot leaf). A one-dimensional network model analogous to an electrical circuit is illustrated in Figure \ref{uniform}, consisting of nodes $i=1,2,\dots,N$. In this example only one xylem conduit is shown as the midline. We assume the water potential in the atmosphere ($\psi_a$) and the baseline potential of leaf water storage ($\psi_s$) are both constant and uniformly distributed along the leaf, and they are more negative than the water potentials in the xylem ($\psi_0,\psi_1,\dots,\psi_N$). Figure \ref{uniform} represents a well-watered leaf, where the base potential $\psi_0$ keeps charging the water-storage capacitors $C_i$, and stomata are wide-open so that water is being released into the atmosphere through transpiration. The resistors $R_i^{(c)}$ represent the resistance in the water-storage pathways, and resistors $R_i^{(a)}$ are the total resistance from xylem to the atmosphere, including outside-xylem resistance for liquid water inside the leaf and stomatal resistance for water vapor. At steady state, the capacitors are fully charged and all $I_i^{(c)}=0$. When stomata are closed, $R_i^{(a)}\rightarrow\infty$ and all $I_i^{(a)}=0$. The summation of all $I_i^{(a)}$ gives the total transpiration current $E=\sum_i I_i^{(a)}$.

The fundamental equations of the electrical analogue are:
\begin{equation}
\label{eqn:kir}
I_{i-1,i}=I_{i,i+1}+I_i^{(a)}+I_i^{(c)}
\end{equation}
\begin{equation}
\psi_{i-1}-\psi_i=R_{i-1,i}I_{i-1,i}
\end{equation}
\begin{equation}
\label{eqn:atm}
\psi_i-\psi_a=R_i^{(a)}I_i^{(a)}
\end{equation}
\begin{equation}
I_i^{(c)}=\frac{\partial}{\partial t}[C_i(\psi_i-\psi_s-R_i^{(c)}I_i^{(c)})].
\end{equation}
At the terminal note (end of the xylem), $I_{N-1,N}=I_N^{(a)}+I_N^{(c)}$. With time-independent $C_i$ and $R_i^{(c)}$, the last equation becomes:
\begin{equation}
\label{eqn:cap}
I_i^{(c)}=C_i\frac{\partial\psi_i}{\partial t}-C_i R_i^{(c)}\frac{\partial I_i^{(c)}}{\partial t}.
\end{equation}
If we also assume the transpiration resistance $R_i^{(a)}$ is time-independent, we can obtain the following equation by substituting Expressions \eqref{eqn:atm} and \eqref{eqn:cap} into the first derivative of Equation \eqref{eqn:kir} with respect to time ($\partial I_{i-1,i}/\partial t=\partial I_{i,i+1}/\partial t+\partial I_i^{(a)}/\partial t+\partial I_i^{(c)}/\partial t$):
\begin{equation}
\label{eqn:org}
\begin{split}
&\quad\frac{\partial I_{i-1,i}}{\partial t}-\frac{\partial I_{i,i+1}}{\partial t}\\
&=\Big(\frac{1}{R_i^{(a)}}+\frac{1}{R_i^{(c)}}\Big)\frac{\partial\psi_i}{\partial t}-\frac{1}{C_i R_i^{(c)}}\Big(I_{i-1,i}-I_{i,i+1}-\frac{\psi_i-\psi_a}{R_i^{(a)}}\Big).
\end{split}
\end{equation}
We will demonstrate how such a spatially uniform, one-dimensional network can be treated as a continuous model when the number of nodes $N$ is large, and can be studied through analytical calculation under certain circumstances. We will also design a numerical method to simulate both the steady state and the time-dependent behavior of a general capacitive network model, which is not necessarily uniform or one-dimensional.

\subsection{Analytical calculation of the one-dimensional model}
\label{sec:model}

We assume the size of a node $l$ is small compared to the total length of the network $L$, so that the number of nodes $N=L/l$ is large. At node $i$ we define a normalized location $x=i/N$ ($\Delta x=l/L$ so that $0\leqslant x\leqslant 1$) which changes continuously, and in the xylem the water potential $\psi$ and current $I$ also change continuously which means $(\psi_{i-1}-\psi_i)/\Delta x\rightarrow -\partial\psi/\partial x$ and $(I_{i-1,i}-I_{i,i+1})/\Delta x\rightarrow -\partial I/\partial x$. In this normalized continuous model, we assume the resistances and capacitances are uniformly distributed and time-independent throughout the network ($R_{i-1,i}=R^{(o)}$, $R_i^{(a)}=R^{(a)}$, $R_i^{(c)}=R^{(c)}$ and $C_i=C^{(o)}$ are all constants), and we define $R=NR^{(o)}$, $R_a=R^{(a)}/N$, $R_c=R^{(c)}/N$ and $C=NC^{(o)}$ as combined parameters for the whole system. Considering $\partial\psi/\partial x=-RI$, we derive the following basic equation from Equation \eqref{eqn:org}:
\begin{equation}
\label{eqn:con}
\frac{\partial^3\psi}{\partial t\partial x^2}=R\Big(\frac{1}{R_a}+\frac{1}{R_c}\Big)\frac{\partial\psi}{\partial t}-\frac{1}{CR_c}\frac{\partial^2\psi}{\partial x^2}+\frac{R}{CR_c R_a}\psi-\frac{R\psi_a}{CR_c R_a}.
\end{equation}

We outline the general time-dependent solution of this equation in Appendix \ref{app:general}. The steady-state average potential in the xylem is:
\begin{equation}
%\label{eqn:barps}
\bar{\psi}=I_0 R_a+\psi_a
\end{equation}
where $I_0=I(x=0)$ is the current entering through the base (see Equation \eqref{eqn:i0} for expression), which is equal to the total transpiration current $E$ at steady state.

For a nontrivial, time-dependent water potential boundary condition, in Appendix \ref{app:tide} we illustrate the solution with an oscillating boundary condition $\psi(x=0,t)=A\cos(\omega_0 t+\varphi)$. We also consider an excised leaf xylem which is initially at the steady state and is cut off at the base at $t=0$, when the base water source is turned off. The calculation is detailed in Appendix \ref{app:cut}. It turns out that both average xylem potential $\bar{\psi}$ and total transpiration current $E$ are exponential decay functions with time at $t>0$, where time constant is $\tau=C(R_c+R_a)$:
\begin{equation}
\label{eqn:pst}
\bar{\psi}(t)=\psi_a+I_0 R_a\exp\Big(-\frac{t}{C(R_c+R_a)}\Big)
\end{equation}
\begin{equation}
\label{eqn:it}
E(t)=I_0\exp\Big(-\frac{t}{C(R_c+R_a)}\Big)
\end{equation}
which means the existence of capacitance $C$ as well as transpiration resistance $R_a$ and water-storage pathway resistance $R_c$ slows down the process of losing water in a drought condition where the leaf loses its water source, illustrating the function of capacitance in a plant's resilience against drought.

\subsection{Numerical simulation of general capacitive networks}
The one-dimensional model in Figure \ref{uniform} can be generalized into two- or higher-dimensional networks, which can include branches and loops. If node $i$ is a node in such a network, and its neighboring nodes are labeled by $n(i)$, we use $I_{n(i),i}$ and $R_{n(i),i}$ to label the current from $n(i)$ to $i$ and the resistance between $n(i)$ and $i$, respectively. Starting from the current relation:
\begin{equation}
\sum_{n(i)}\frac{\partial I_{n(i),i}}{\partial t}=\frac{\partial I_i^{(a)}}{\partial t}+\frac{\partial I_i^{(c)}}{\partial t},
\end{equation}
we derive the water potential relation at $i$:
\begin{equation}
\begin{split}
&\quad\Big(\sum_{n(i)}\frac{1}{R_{n(i),i}}+\frac{1}{R_i^{(a)}}+\frac{1}{R_i^{(c)}}\Big)\frac{\partial\psi_i}{\partial t}-\sum_{n(i)}\frac{1}{R_{n(i),i}}\frac{\partial\psi_{n(i)}}{\partial t}\\
&=\frac{1}{C_i R_i^{(c)}}\Big(\frac{\psi_a-\psi_i}{R_i^{(a)}}+\sum_{n(i)}\frac{\psi_{n(i)}-\psi_i}{R_{n(i),i}}\Big)
\end{split}
\end{equation}
where $\psi_{n(i)}$ is the water potential at node $n(i)$. The problem of solving the time-dependent behavior of $\psi_i$ is organized into a matrix equation $\vek{A}\vek{x}=\vek{b}$, where at time $t$, the vector to be solved is:
\begin{equation}
\vek{x}=\left(\frac{\partial\psi_1}{\partial t},\frac{\partial\psi_2}{\partial t},\dots,\frac{\partial\psi_i}{\partial t},\dots,\frac{\partial\psi_N}{\partial t}\right)^{\text{T}}
\end{equation}
which contains the time derivatives of water potential at all nodes. The $i$th element of vector $\vek{b}$ is:
\begin{equation}
\vek{b}_i=\frac{1}{C_i R_i^{(c)}}\Big(\frac{\psi_a-\psi_i(t)}{R_i^{(a)}}+\sum_{n(i)}\frac{\psi_{n(i)}(t)-\psi_i(t)}{R_{n(i),i}}\Big)
\end{equation}
and the elements in the invertible and symmetric matrix $\vek{A}$ are determined by:
\begin{widetext}
\begin{equation}
\vek{A}_{i,j}=\left\{\begin{array}{lr}\sum_{n(i)}1/R_{n(i),i}+1/R_i^{(a)}+1/R_i^{(c)}&i=j\\ -1/R_{j,i}&j\text{ is neighbor of }i\\ 0&i\neq j\text{ \& }j\text{ is not neighbor of }i\end{array}\right..
\end{equation}
\end{widetext}
If the node $i$ is connected to one or more water sources, the external water potentials must be included in the evaluation of $\vek{b}_i$ and $\vek{A}_{i,j}$. For example, if node $i$ is connected to a base potential $\psi_p$, we have $\vek{b}_i=1/(C_i R_i^{(c)})[(\psi_p-\psi_i)/R_{p,i}+(\psi_a-\psi_i)/R_i^{(a)}+\sum_{n(i)}(\psi_{n(i)}-\psi_i)/R_{n(i),i}]$ and $\vek{A}_{i,i}=1/R_{p,i}+\sum_{n(i)}1/R_{n(i),i}+1/R_i^{(a)}+1/R_i^{(c)}$, where $R_{p,i}$ is the resistance between $i$ and base (location of water source).

To simulate the dynamics of the network, we start from an initial water status $\psi_i(t=0)$, calculate $\vek{b}$ and then $\vek{x}=\vek{A}^{-1}\vek{b}$ at the current time $t$, and update the water potentials after a small time step $\Delta t$:
\begin{equation}
\psi_i(t+\Delta t)=\psi_i(t)+\frac{\partial\psi_i}{\partial t}\Delta t.
\end{equation}
To numerically calculate the steady state where all $\partial\psi_i/\partial t=0$, we organize the equation $(\psi_a-\psi_i)/R_i^{(a)}+\sum_{n(i)}(\psi_{n(i)}-\psi_i)/R_{n(i),i}=0$ into another matrix equation $\vek{B}\vek{y}=\vek{a}$, where $\vek{y}_i=\psi_i$, $\vek{a}_i=\psi_a/R_i^{(a)}$ (or $\vek{a}_i=\psi_p/R_{p,i}+\psi_a/R_i^{(a)}$ if $i$ is connected to a water source $\psi_p$), and $\vek{B}_{i,j}=\vek{A}_{i,j}$ if $i\neq j$ while $\vek{B}_{i,i}=\sum_{n(i)}1/R_{n(i),i}+1/R_i^{(a)}$ (or $\vek{B}_{i,i}=1/R_{p,i}+\sum_{n(i)}1/R_{n(i),i}+1/R_i^{(a)}$). By solving $\vek{y}=\vek{B}^{-1}\vek{a}$ we can calculate the steady-state water potentials in the network.

\subsection{The lumped model and its parameter selection}
\label{sec:lump}
\begin{figure}[hbt!]
\centering
\includegraphics[width=0.3\textwidth]{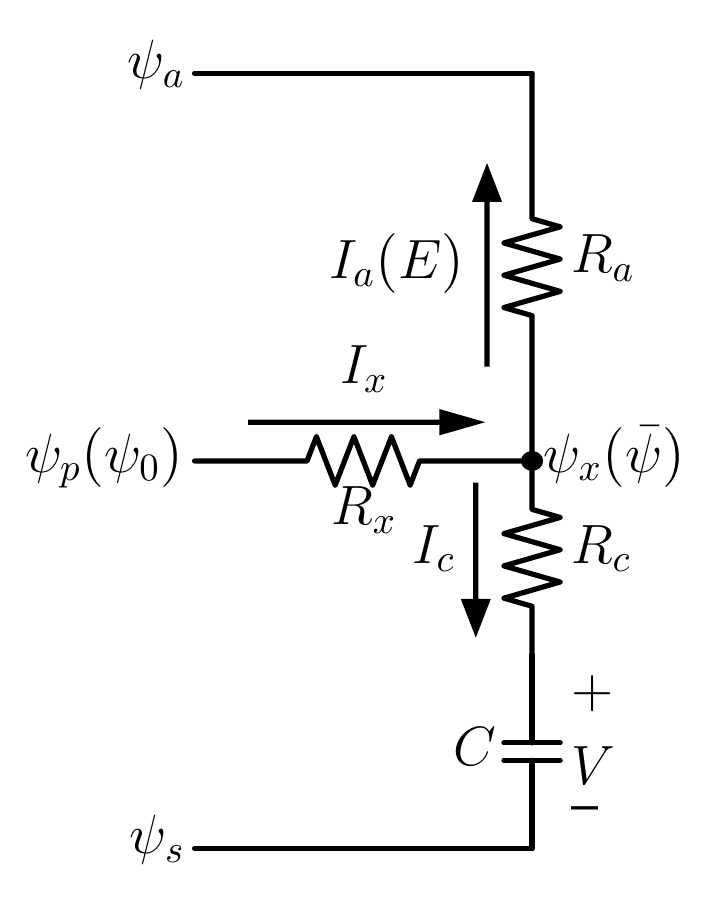}
\caption{A lumped-element model which is approximately equivalent to the one-dimensional network in Figure \ref{uniform}. Here $\psi_p=\psi_0$ is the water potential at the base, and $\psi_x=\bar{\psi}$ is the average potential in the xylem. Currents $I_a=E$ and $I_c$ are total transpiration current and capacitor charging current, respectively.}
\label{lump}
\end{figure}

In this subsection we introduce a highly lumped-element model, which is shown in Figure \ref{lump}. We compare the basic functions of the one-dimensional model in Figure \ref{uniform} (left) with those of Figure \ref{lump} (right):
\begin{align}
&I_{0,1}=\sum_i I_i^{(a)}+\sum_i I_i^{(c)}& &I_x=I_a+I_c\label{comp1}\\
&\psi_0-\psi_N=\sum_i R_{i-1,i}I_{i-1,i}& &\psi_p-\psi_x=R_x I_x\label{comp2}\\
&\bar{\psi}-\psi_a=\frac{\sum_i R_i^{(a)}I_i^{(a)}}{N}& &\psi_x-\psi_a=R_a I_a\label{comp3}\\
&\bar{\psi}-\bar{V}-\psi_s=\frac{\sum_i R_i^{(c)}I_i^{(c)}}{N}& &\psi_x-V-\psi_s=R_c I_c\label{comp4}
\end{align}
where $\bar{\psi}=\sum_i\psi_i/N$ and $\bar{V}=\sum_i V_i/N$ are average xylem water potential and average capacitor voltage, respectively. We investigate the equivalence of the two sets of equations by defining $I_x=I_{0,1}$, $I_a=\sum_i I_i^{(a)}$, $I_c=\sum_i I_i^{(c)}$, $\psi_x=\bar{\psi}$ and $V=\bar{V}$. In Subsection \ref{sec:model}, for a uniform one-dimensional network where circuit elements are evenly distributed, we have defined $R_a$ and $R_c$ as grouped elements for the whole system, and these definitions will make the equations in \eqref{comp1}, \eqref{comp3} and \eqref{comp4} equivalent. To make equations in \eqref{comp2} also approximately equivalent, we can define $R_x=\sum_{i}R_{i-1,i}/3$, which means $R_x=R/3$ for a uniform network in Subsection \ref{sec:model}. See Appendix \ref{app:lump} for the details of this reasoning.

The lumped model in Figure \ref{lump} is described by the equation:
\begin{equation}
(\frac{1}{R_a}+\frac{1}{R_x}+\frac{1}{R_c})\frac{\partial\psi_x}{\partial t}=-\frac{1}{CR_c}(\frac{1}{R_a}+\frac{1}{R_x})\psi_x+\frac{1}{CR_c}(\frac{\psi_a}{R_a}+\frac{\psi_p}{R_x})
\end{equation}
which gives rise to:
\begin{equation}
\psi_x=\frac{\frac{\psi_a}{R_a}+\frac{\psi_p}{R_x}}{\frac{1}{R_a}+\frac{1}{R_x}}+D\exp\left(-\frac{\frac{1}{CR_c}(\frac{1}{R_a}+\frac{1}{R_x})}{\frac{1}{R_a}+\frac{1}{R_x}+\frac{1}{R_c}}t\right)
\end{equation}
where the first part is the steady state and the second part is the dynamics, where the parameter $D$ is the difference between the initial state and the steady state. The time constant becomes $\tau=C(R_c+R_a)$ if the model describes a leaf removed from plant ($R_x\rightarrow\infty$), the same result as in the uniform network model (Subsection \ref{sec:model}).

The lumped model in Figure \ref{lump} is helpful for estimating the importance of the stomatal sensitivity to leaf water content (contained by capacitors) in controlling the transpiration rate. We suppose the water content of a well-watered leaf is $W_0$ and that when the leaf is slightly dehydrated to a water content $W<W_0$, both whole-leaf lumped capacitor and stomatal resistance depend linearly on $W$ (expanded to the first order). Thus we assume $V(W)=W/C$ and $R_a(W)=r_a+s(W_0-W)$, where $C$ is a regular capacitance and $s$ is a linear measure of the sensitivity of $R_a$ (whose minimum value is $r_a$) to $W$, so that $R_a$ increases with decreasing $W$ according to the expected behavior of stomata. From Equations \eqref{comp1}--\eqref{comp4}, we calculate the expressions of $\psi_x$, $I_x$, $I_a$ and $I_c$ in terms of given terminal water potentials and electrical traits including $V(W)$ and $R_a(W)$. From the dependence of $I_c$ on $W$ and the relation $I_c=\dif W/\dif t$, we calculate a steady-state expression for $W$ which represents the static water status of a living leaf. Furthermore, by substituting this steady-state $W$ into the expression of $I_a$, we obtain the following estimate of the relationship between terminal potentials and $I_a$ (showing the highest-order term):
\begin{equation}
\psi_p-\psi_a\approx CR_x s\cdot I_a^2+\cdots
\end{equation}
which emphasizes the essential functions of leaf capacitance $C$, xylem hydraulic resistance $R_x$ and the stomatal sensitivity to water content in limiting the increase of transpiration current $I_a$ with an increasing water potential deficit $\psi_p-\psi_a$ between inside and outside of the leaf. In this way both capacitance and stomatal control are shown to be helpful for keeping water content and reducing water loss in transpiration.

\section{Computation and simulation results of uniform model leaf}
\label{sec:result}

\begin{figure}[hbt!]
\centering
\includegraphics[width=0.5\textwidth]{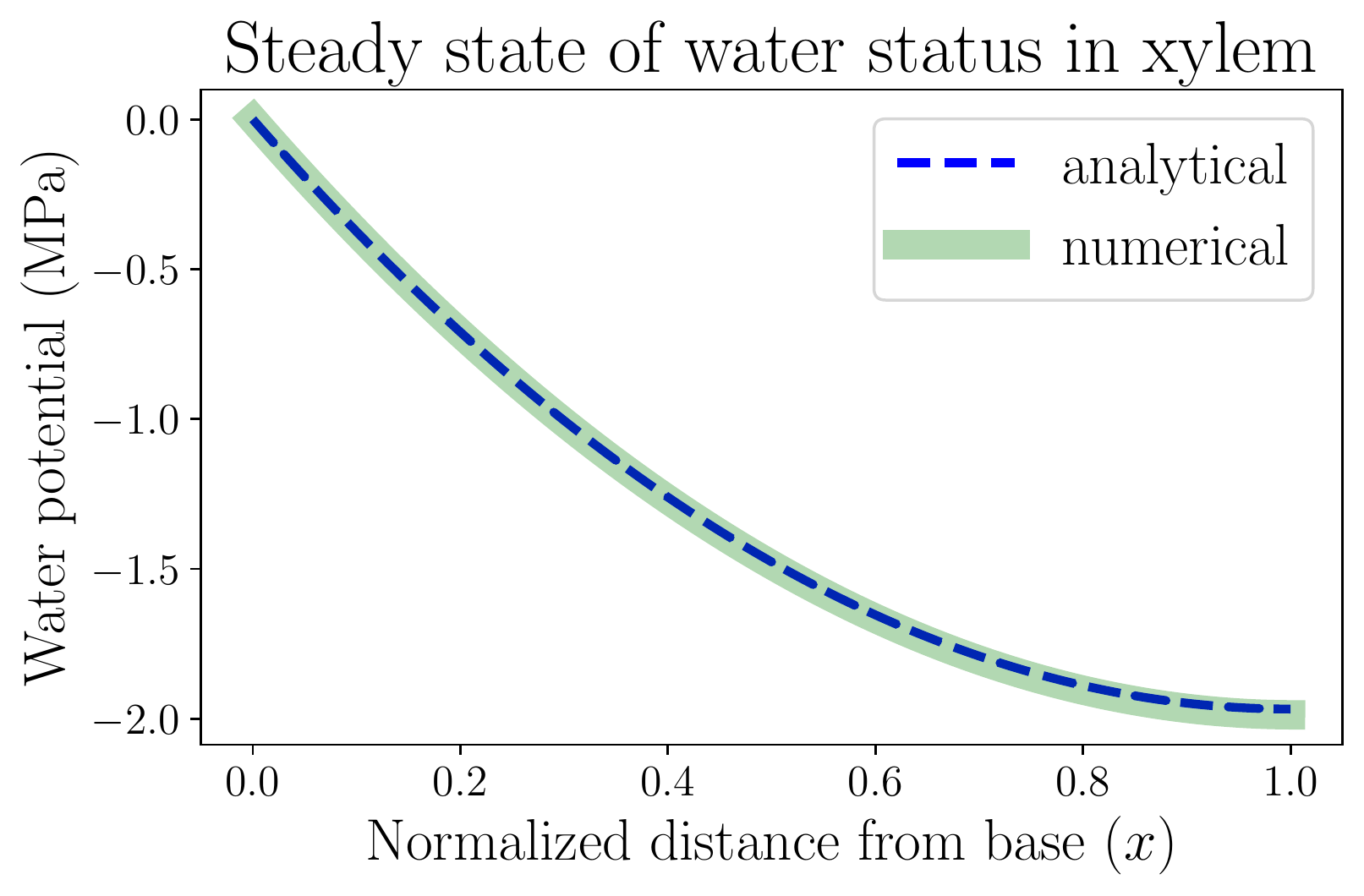}
\caption{The spatially dependent water potential in a living, uniform model leaf at the steady state. The water potential decreases from 0 ($x=0$, the leaf base) to about \SI{-2}{\mega\pascal} ($x=1$, the leaf tip). The one-dimensional numerical simulation result follows the analytical expression. Parameters used in the modeling include: $\psi_a=\SI{-100}{\mega\pascal}$, $R=\SI{2}{\mega\pascal}\cdot\SI{}{\m}^2\cdot\SI{}{\s}\cdot\SI{}{\milli\mol}^{-1}$, and $R_a=\SI{50}{\mega\pascal}\cdot\SI{}{\m}^2\cdot\SI{}{\s}\cdot\SI{}{\milli\mol}^{-1}$. The steady-state transpiration rate of this model leaf is $E=\SI{1.97}{\milli\mol}\cdot\SI{}{\m}^{-2}\cdot\SI{}{\s}^{-1}$.}
\label{ss}
\end{figure}

In this section we apply the analytical calculation (for one-dimensional continuous models) and numerical simulation methods (for discretized network models) introduced in the previous section to the study of hydraulic behaviors of a model leaf, by making use of biologically relevant parameters. One of these results is the spatially dependent water potential profile at the steady state of a living uniform leaf shown in Figure \ref{ss}. We assume the plant is well watered and estimate that the water potential at the leaf base $\psi(x=0)$ is approximately $\psi_0=0$ where $x$ is the normalized distance from base ($x=0$) to tip ($x=1$). In order to generate a clear spatial pattern of water status with significant spatial variations, we select an atmospheric water potential $\psi_a=\SI{-100}{\mega\pascal}$, which represents moderately dry air conditions outside the leaf stomata. From this $\psi_a$ value, we can estimate the relative humidity (RH) in the air through the relationship:
\begin{equation}
\label{eqn:vpd}
\psi_a=\frac{\bar{R}T}{v}\ln\left(\frac{\text{RH}}{100\%}\right)
\end{equation}
with ideal gas constant $\bar{R}=\SI{8.3145}{J}\cdot\SI{}{\mol}^{-1}\cdot\SI{}{\kelvin}^{-1}$, room temperature $T=\SI{298.15}{\kelvin}$, and liquid water molar volume $v\approx\SI{18}{\milli\liter}/\SI{}{\mol}$ \cite{Buckley2019}. The estimated relative humidity $\text{RH}\approx48\%$ can be used to calculate the vapor pressure deficit (VPD) between the inner air space of a leaf and the outside atmosphere (across stomata), which is estimated to be $\text{VPD}=(1-\text{RH})P\approx\SI{1.64}{\kilo\pascal}$, where $P=\SI{3.17}{\kilo\pascal}$ is the saturation vapor pressure of water ($\text{RH}=100\%$) at room temperature. The concept of VPD is usually used in plant biology as the driving force of transpiration.

The resulting steady-state water potential profile illustrates a monotonically decreasing trend from base to tip, so that the water flow is unidirectional in this uniform model leaf. Leaf water potential is maintained above $\SI{-2}{\mega\pascal}$, below which xylem conduits may embolize and flow may start to cease. We use hydraulic resistance parameters including xylem total resistance $R=\SI{2}{\mega\pascal}\cdot\SI{}{\m}^2\cdot\SI{}{\s}\cdot\SI{}{\milli\mol}^{-1}$ and resistance from xylem to air $R_a=\SI{50}{\mega\pascal}\cdot\SI{}{\m}^2\cdot\SI{}{\s}\cdot\SI{}{\milli\mol}^{-1}$ (so that xylem conductance is 25 times the conductance into air). This large resistance $R_a$ (and low conductance) in the pathway from xylem to the atmosphere is mainly comprised of two parts, namely the stomatal resistance $R_s$ and the outside-xylem resistance $R_{ox}$, which is between xylem and stomata. If we consider $R_a$ as a variable dependent on water content $W$, for example in the lumped-element model introduced in the previous section, we can assume $R_a(W)=R_s(W)+R_{ox}$ in which $R_s(W_0)=0$ and $R_a(W_0)=R_{ox}$ when the leaf is fully hydrated ($W=W_0$) and stomata are open, and thus the outside-xylem resistance through leaf tissue (such as mesophyll) provides the minimum resistance in the transpiration pathway.

The analytical solution in Figure \ref{ss} is obtained from Equation \eqref{eqn:ss}, which is a monotonically decreasing function. The numerical simulation of network is conducted by discretizing the leaf into 100 nodes from base to tip, each node with parameters $R^{(o)}=\SI{0.02}{\mega\pascal}\cdot\SI{}{\m}^2\cdot\SI{}{\s}\cdot\SI{}{\milli\mol}^{-1}$ and $R^{(a)}=\SI{5000}{\mega\pascal}\cdot\SI{}{\m}^2\cdot\SI{}{\s}\cdot\SI{}{\milli\mol}^{-1}$. The simulation result reproduces the spatial distribution of water potential described by the analytical solution, also validating the usefulness of the simulation method which can be used for a more general, expanded network model. With these parameters, the average xylem water potential and total transpiration rate are calculated as $\bar{\psi}=\SI{-1.31}{\mega\pascal}$ and $E=I_0=\SI{1.97}{\milli\mol}\cdot\SI{}{\m}^{-2}\cdot\SI{}{\s}^{-1}$ according to Equations \eqref{eqn:barps} and \eqref{eqn:i0}, which are of the same order of magnitude as typical experimental values (which can be found in textbooks like \cite{Jones2013}). If the distribution of water-storage capacitance is also uniform along the leaf, the local leaf water content would taper toward the tip, which can be tested by experiments (see Section \ref{sec:disc}).

\subsection{Exactly solvable model: dehydration of an excised leaf}

\begin{figure*}[hbt!]
\centering
\includegraphics[width=0.8\textwidth]{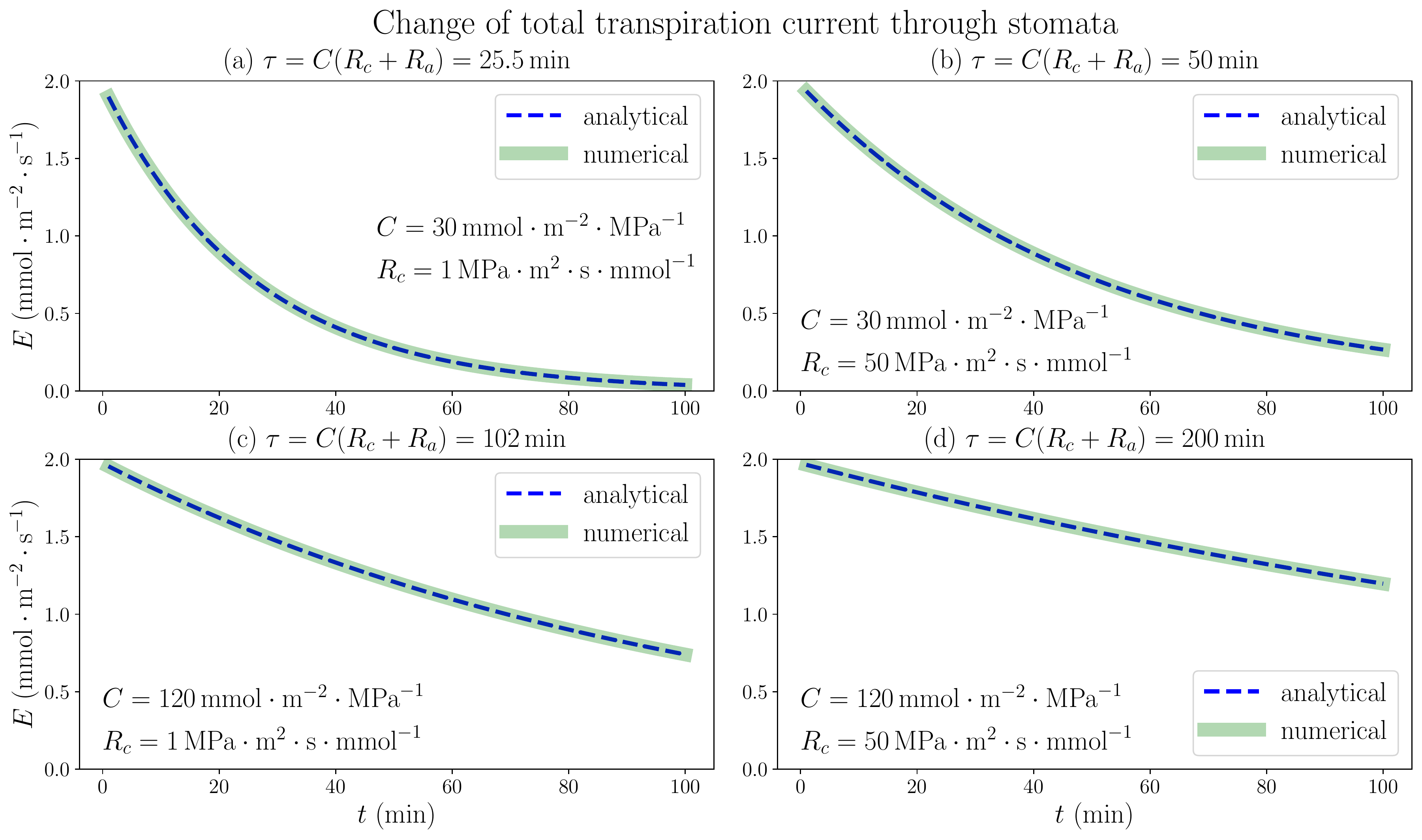}
\caption{The time dependence of total transpiration rate through stomata when a living, uniform model leaf in steady state is removed from a plant and the water source is closed at the base. The 1D simulation results match the analytical exponential decay expressions, with time constants $\tau=C(R_c+R_a)$. The temporal variations of average water potential in xylem are similar. Various sets of whole-leaf capacitance $C$ and xylem-to-capacitor resistance $R_c$ values are used, while other parameters are the same as those in Figure \ref{ss}. The effect of increasing (or decreasing) the capacitance $C$ on the rate of change is shown to be much larger than that of the increase (or decrease) of $R_c$.}
\label{tt}
\end{figure*}

We consider the analytically solved situation of excising a living uniform model leaf in steady state from plant (see Appendix \ref{app:cut}). The results in Equations \eqref{eqn:pst} and \eqref{eqn:it} show that the time dependence of both average xylem water potential and total transpiration current is characterized by a time constant $\tau=C(R_c+R_a)$ in an exponentially decaying trend in the dehydration process, as long as the stomatal resistance is kept unchanged. Figure \ref{tt} illustrates the time dependence of total transpiration rate $E$, which continuously decreases from the steady-state value $\SI{1.97}{\milli\mol}\cdot\SI{}{\m}^{-2}\cdot\SI{}{\s}^{-1}$ instead of going through a drastic change because of the existence of capacitance. In addition to the same parameters $\psi_0$, $\psi_a$, $R$ and $R_a$ used in Figure \ref{ss}, we use different sets of capacitance value $C$ and resistance $R_c$ from xylem to capacitor for the whole leaf as labeled in each subplot of Figure \ref{tt}. The discretized numerical results, which are obtained by conducting the simulation in a 100-node network where $C^{(o)}=C/100$ and $R^{(c)}=100R_c$ at each node with time interval $\Delta t=\SI{0.01}{\min}$, match the analytical expressions with time constant $\tau$, revalidating the simulation method. These results are based on the critical assumption that all hydraulic elements including all resistances are constant, which leads to a large drop of average xylem water potential $\bar{\psi}$ to very negative, non-physiological values in a matter of minutes (see supplementary information). When $\bar{\psi}$ drops to nearly $\psi_a$, the transpiration rate would become nearly 0, as predicted by Equations \eqref{eqn:pst} and \eqref{eqn:it}. The calculations shown here are mainly used to explore the function of capacitors in the adjustment of leaf water status, emphasizing their importance for the stabilization and resilience of plant hydraulics.

As predicted by the theoretical calculations and proven by the numerical simulations, both capacitance $C$ and resistance $R_c$ are shown to play an important role in the dehydration dynamics of a model leaf experiencing the removal of water source, which models the behavior of plant water status in a severe drought condition. It turns out that within the ranges of parameters we choose, the variation of $C$, which is directly proportional to $\tau$, would exert a much larger influence on the time scale and rate of change in the dehydration process than the effect of varying $R_c$. While $R_c$ increases from $\SI{1}{\mega\pascal}\cdot\SI{}{\m}^2\cdot\SI{}{\s}\cdot\SI{}{\milli\mol}^{-1}$ (same order of magnitude as $R$) to $\SI{50}{\mega\pascal}\cdot\SI{}{\m}^2\cdot\SI{}{\s}\cdot\SI{}{\milli\mol}^{-1}$ (comparable to $R_a$), which is a fifty times increase, the time constant $\tau$ is only increased by less than two times. The resulting time constants (tens of minutes) and total changes of transpiration rate ($\sim\SI{1}{\milli\mol}\cdot\SI{}{\m}^{-2}\cdot\SI{}{\s}^{-1}$) in \SI{100}{\min} are comparable to the magnitudes of corresponding experimental measurement results, verifying our selection of hydraulic parameters. These observations of a model leaf indicate that an effective strategy for a plant to be more resilient under water stress and to survive a drought would be to enlarge its water-storage capacitance, the ability to contain large amount of water, rather than to increase the resistance of pathways connecting xylem and capacitors. In a real-life plant, whose stomatal resistance is changeable and sensitive to the water status, a drought stressed condition and decreasing water content would trigger the closing of stomata, drastically increasing $R_a$, which also prolongs the time constant and slows down the decrease of water potential (while ceasing transpiration), providing another effective strategy to overcome water stress.

\subsection{Numerical simulations of leaf water status in changing environments}

\begin{figure*}[hbt!]
\centering
\includegraphics[width=0.8\textwidth]{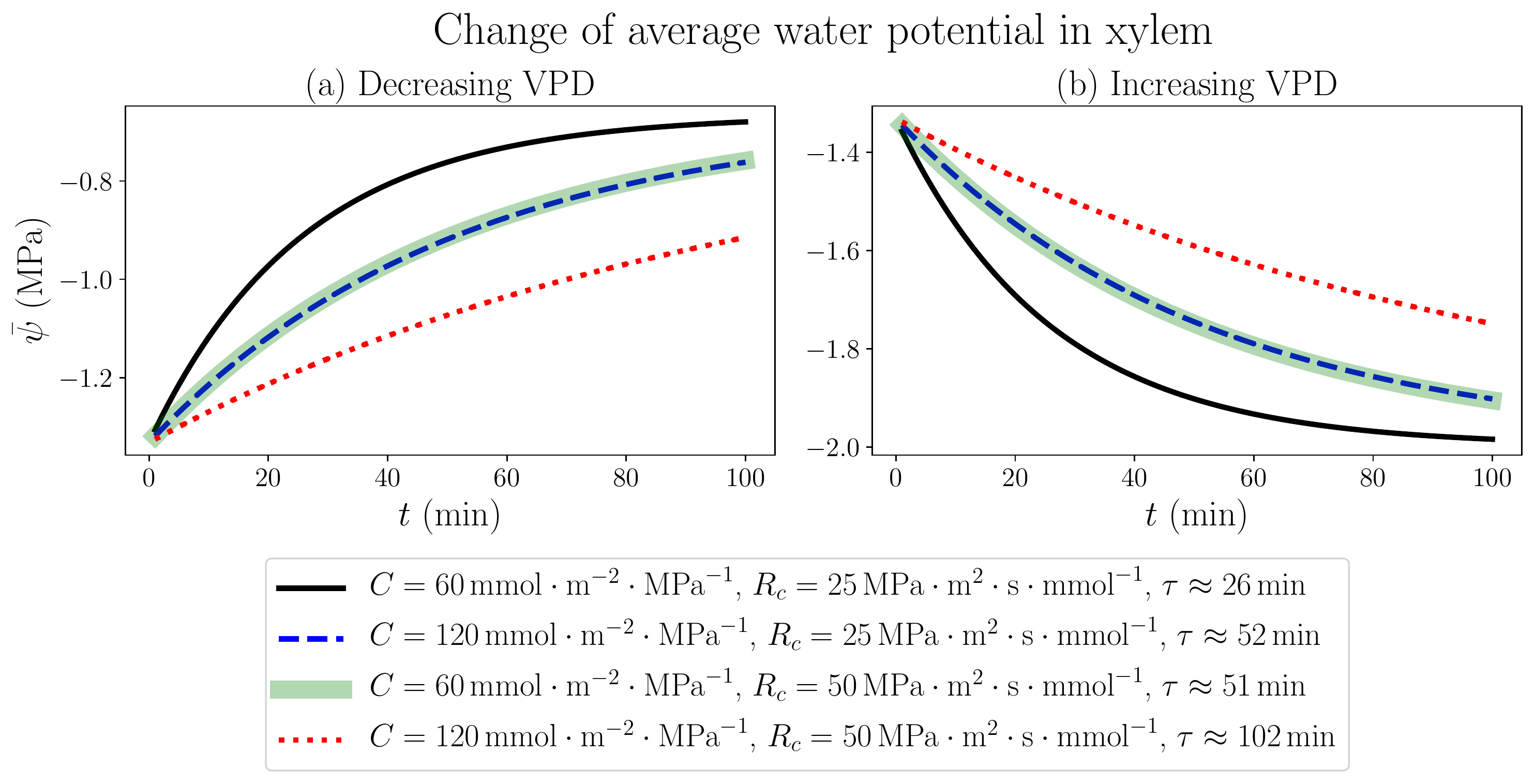}
\caption{The time dependence of average water potential in xylem, when $\psi_a$ is abruptly increased from \SI{-100}{\mega\pascal} to \SI{-50}{\mega\pascal} (a) where the air is wetted and VPD is decreased from \SI{1.64}{\kilo\pascal} to \SI{0.965}{\kilo\pascal}, or decreased from \SI{-100}{\mega\pascal} to \SI{-150}{\mega\pascal} (b) where the air is dried and VPD is increased from \SI{1.64}{\kilo\pascal} to about \SI{2.1}{\kilo\pascal}. Various sets of $C$ and $R_c$ values are used, while other parameters are the same as those in Figure \ref{ss}. In both situations, the time constants $\tau$ corresponding to the same $C$ and $R_c$ values are identical as labeled in the legend.}
\label{vpd}
\end{figure*}

We consider the response of a living, fully hydrated plant leaf to an instant change of atmospheric conditions, such as a sudden increase or decrease of relative humidity (RH) in the air, which is related to a rise or drop of atmospheric water potential through Equation \eqref{eqn:vpd}. The results are generated by simulating the 100-node discretized network (with simulation time step $\Delta t=\SI{0.01}{\min}$). We start from modeling a well-watered leaf in the steady state using the parameters for Figure \ref{ss} ($\psi_a=\SI{-100}{\mega\pascal}$, $\text{RH}\approx48\%$ and $\text{VPD}\approx\SI{1.64}{\kilo\pascal}$), which is the same initial state in Figure \ref{tt}, and at time $t=0$ instantly change the value of $\psi_a$, resulting in a continuous change of average water potential in xylem starting from \SI{-1.31}{\mega\pascal} as shown in Figure \ref{vpd}. As $\psi_a$ is raised to \SI{-50}{\mega\pascal}, when the air is more humid ($\text{RH}\approx70\%$) and the vapor pressure deficit across stomata becomes smaller ($\text{VPD}\approx\SI{0.965}{\kilo\pascal}$), $\bar{\psi}$ also increases gradually toward a new steady-state value \SI{-0.656}{\mega\pascal}, which is determined by the new $\psi_a$ value through Equations \eqref{eqn:barps} and \eqref{eqn:i0} with constant resistance values $R$ and $R_a$. Similarly, as $\psi_a$ is dropped to \SI{-150}{\mega\pascal}, when the air is drier ($\text{RH}\approx34\%$) and VPD becomes larger (\SI{2.1}{\kilo\pascal}), $\bar{\psi}$ will decrease with time to an ultimate steady-state value \SI{-1.97}{\mega\pascal}. The analytical expressions for the time dependence of $\bar{\psi}$, as illustrated in Figure \ref{vpd} (a) and (b) for increasing and decreasing $\psi_a$ respectively, are not explicitly available, but we can fit the profiles of $\bar{\psi}$ to exponential decay curves with time constant $\tau$ shown in the legend. The time constant for a certain set of $C$ and $R_c$ values turns out to be identical in both air wetting and drying situations, proving that the specific dynamics of leaf water status depends only on the internal hydraulic traits rather than external environments. The total transpiration rate $E$ obtained in this modeling changes with time in a slightly different way, in which $E$ would abruptly jump to a new lower (increasing $\psi_a$) or higher value (decreasing $\psi_a$), and then gradually change with the same exponentially decaying trend and time constant as $\bar{\psi}$, to steady-state values $\SI{0.987}{\milli\mol}\cdot\SI{}{\m}^{-2}\cdot\SI{}{\s}^{-1}$ and $\SI{2.96}{\milli\mol}\cdot\SI{}{\m}^{-2}\cdot\SI{}{\s}^{-1}$ for cases in (a) and (b) respectively (see supplementary information).

The continuous change of average xylem water potential and the avoidance of drastic variation in a short time are another illustration of the functions of leaf water-storage capacitance $C$, as well as resistance $R_c$ from xylem to capacitors, in stabilizing the plant hydraulics and reducing the variation of water content. The instant changes of $\psi_a$ and VPD are used to model the effects of transient changes of wind condition, which would cause the leaf to lose water and dehydrate under the influence of a drying atmosphere even when the plant is well watered. It turns out that the effects of changing $C$ and changing $R_c$ on the time dependence of $\bar{\psi}$ are similar in this case within the ranges of parameters we choose. While making the capacitance twice as large will exactly increase the time constant to two times, suggesting a direct proportionality between $C$ and $\tau$, we find that doubling $R_c$ will also make $\tau$ increase to a little lower than two folds. The weaker effect of $R_c$ is possibly related to the presence of unchanging $R_a$. This observation once again points to the effective strategy for a plant to overcome hydraulic destabilization and water loss due to negative environmental disturbances, by increasing either $C$ or $R_c$ of the leaf.

\begin{figure*}[hbt!]
\centering
\includegraphics[width=0.6\textwidth]{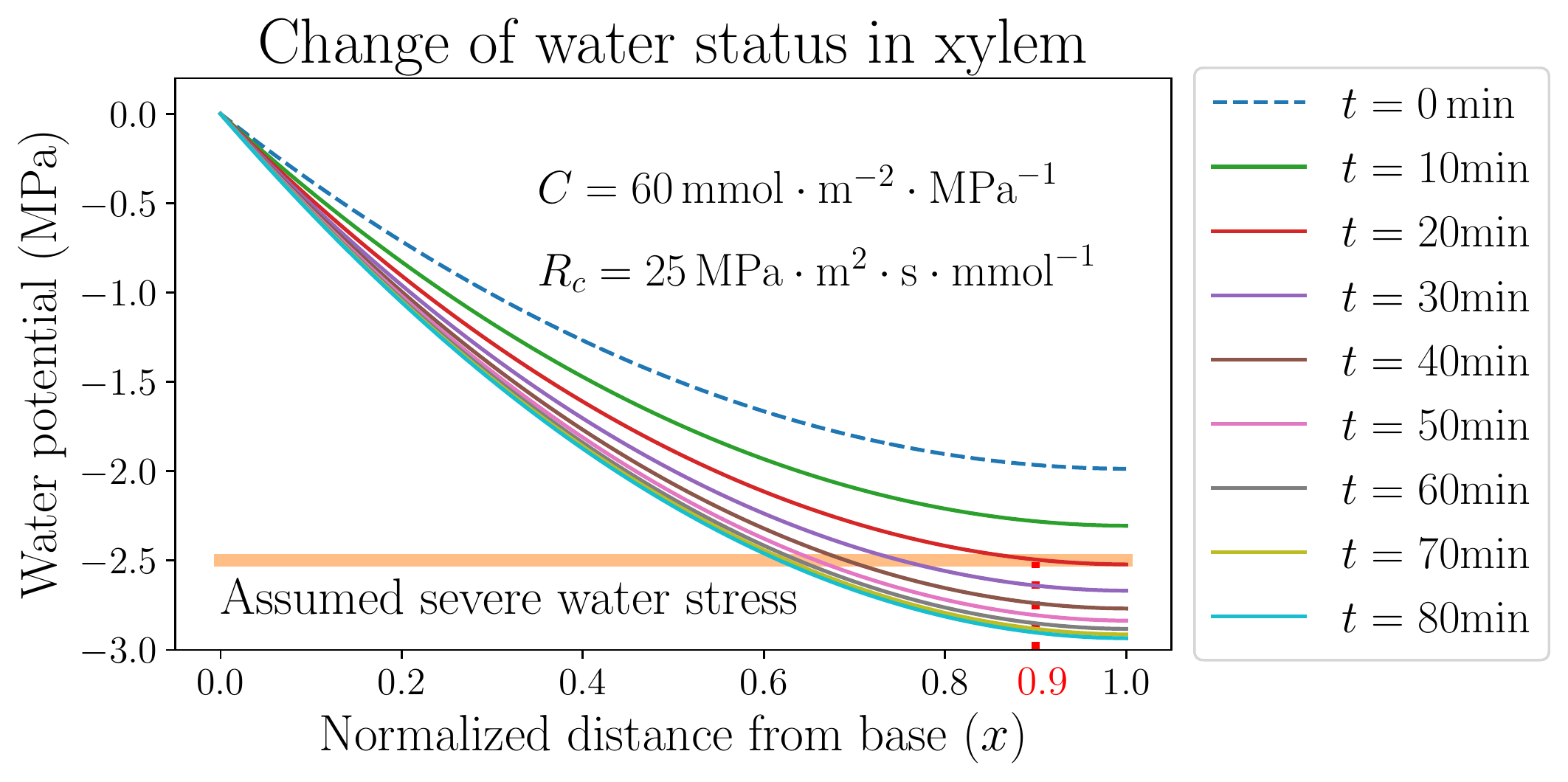}
\caption{The temporal variation of spatially dependent water potential in xylem, when the atmospheric water potential $\psi_a$ is instantly decreased from \SI{-100}{\mega\pascal} to \SI{-150}{\mega\pascal} where the air is dried. Parameters used are identical to those used for the black curves in Figure \ref{vpd}. The horizontal line labels the upper limit of presumed water potential under severe stress.}
\label{temp}
\end{figure*}

The stabilization of $\bar{\psi}$, however, only represents an average effect over the whole leaf from base to tip. The local xylem water potential $\psi$, which varies spatially in the leaf, would stay close to 0 near the base ($x=0$) but would still become very negative near the tip ($x=1$). An example is demonstrated in Figure \ref{temp}, where the dynamics of spatially dependent water potential is simulated and its snapshots are plotted at multiple time points along the simulation. The figure shows the detailed changes of $\psi$ according to the same hydraulic parameters and atmospheric condition of the black solid line in Figure \ref{vpd} (b), where $\psi_a$ decreases from \SI{-100}{\mega\pascal} to \SI{-150}{\mega\pascal} instantly at $t=0$ and VPD increases from \SI{1.64}{\kilo\pascal} to about \SI{2.1}{\kilo\pascal}. Within tens of minutes, the average potential $\bar{\psi}$ is stabilized at around \SI{-2}{\mega\pascal} (specifically \SI{-1.97}{\mega\pascal}) which is the new steady state at the lower $\psi_a$, while the profile of $\psi$ maintains a monotonically declining trend but lowers quantitatively with time. The rate of the lowering of $\psi$ is initially faster and slows down later, corresponding to the exponential decay of $\bar{\psi}$. Even though the xylem average water potential is always higher than \SI{-2}{\mega\pascal}, the local water potential near the tip would experience severe water stress (which is assumed to be lower than \SI{-2.5}{\mega\pascal} here) after a short time. This severely stressed water potential is presumably lower than the value required for the normal functioning of a plant leaf, and would thoroughly dehydrate the leaf portion under this negative potential, making it lose physiological functions. For example, in about \SI{20}{\min} after the instant change of $\psi_a$ and VPD, the leaf portion at $x>0.9$ (between the vertical dotted line and the tip at $x=1.0$ in Figure \ref{temp}) would experience the low water potential and severe stress, and would be quickly dehydrated even when the average water status of the whole leaf is still relatively high. As the simulation proceeds with time, the leaf portion near the tip undergoing severe water stress would enlarge and the left boundary of this region (the vertical dotted line) would move to smaller $x$. This discovery calls for caution when studying the average water status of a leaf, which may be in a safe range for the leaf tissue to stay healthy, while the local water potential (especially at the tip) may be very negative and the leaf can be partly dehydrated, losing part of its functionality. A large capacitance $C$, and also large resistances $R_c$ and $R_a$, could help to delay the lowering of water potential by increasing the time scale $\tau$, so that even the tip potential could be held at a high level to avoid dehydration for a relatively long time. A real-life living plant leaf would most likely close stomata and immediately raise $R_a$, when facing dry wind in the air and going through quickly rising VPD.

\section{Discussion}
\label{sec:disc}

One of the central assumptions of our theoretical modeling work on uniform grass leaf model is the constant resistance $R_a$ from xylem to the atmosphere. This assumption implies that the stomatal resistance to water vapor flow is steady and independent of environmental changes within the modeled time period, a highly hypothetical stomatal behavior which is usually not the case in a living plant but is most helpful for focusing on the effects of water-storage capacitance. This assumption is not trivial or baseless even from a plant biological point of view. It is shown that beginning with a well-watered state, both stomatal conductance and transpiration rate are stabilized and would not decrease significantly even when leaf water potential starts to drop, as long as the potential is higher than a threshold that causes stomata to react, increasing their resistance and ultimately closing \cite{Brodribb2009,Choat2018}. In fact, the dynamic processes modeled in this work all start from a well-watered steady state, and the results are meaningful for the study of initial changes and reactions of a leaf blade in response to instant or short-time variations of water conditions, which are exactly what is considered in the hypothesized dynamic scenarios. In order to model the long-term dehydration dynamics, we would need to incorporate the dependence of stomatal resistance (and thus $R_a$) on the leaf water status, which is briefly discussed in terms of lumped-element model where $R_a$ is assumed to be a linear function of water content, and also in Section \ref{sec:result} where $R_a$ is separable into stomatal resistance $R_s$ (water content dependent) and outside-xylem $R_{ox}$ (independent of water content). If the dehydration lasts longer and the leaf approaches severe stress (such as water potential below the limit of severe water tress in Figure \ref{temp}), the xylem hydraulic conductance would also start to decrease due to the formation of embolism or cavitation, air bubbles blocking water flow in xylem vessels \cite{Sack2006,Stroock2014,Tyree1991,Jones2013}. The dependence of xylem conductance on water potential is conventionally approximated as a logistic function with the shape of a sigmoid curve, in which the loss of conductance is negligible at relatively high potential and grows more rapidly with further lowering potential. Indeed, the incorporation of xylem and stomatal conductances dependent on water potential is applied in several theoretical models, in which the stomatal dependence is also treated as sigmoidal or approximately piecewise linear functions \cite{Mencuccini2019}. Most recently, the sigmoidal behaviors of both xylem and stomatal conductances are applied in a spatially explicit study \cite{Jain2021}. The specific biophysical and biochemical mechanisms that control stomatal opening through water status, including the turgor pressure of guard cells and epidermal cells around stomata and the use of plant hormone abscisic acid, are broadly explored by existing literature \cite{Buckley2005}. If the spatial variation of stomatal conductance (or resistance) is also included, the model would need to take into account anatomical data over leaf blade, including sizes and spatial distributions of stomata and xylem \cite{Ocheltree2012,Rockwell2017}. Our model does not address leaf anatomy directly, but can easily incorporate the quantitative dependences and relationships between local resistances and water potential or content in simulations.

Under the assumption of constant stomatal opening, our computation and simulation results indicate that both capacitance $C$ and the resistance $R_c$ from xylem to capacitors play significant roles in defining the time constant $\tau$, which determines the rate of change in a dehydration (or hydration) dynamics. Both $C$ and $R_c$ are positively related to $\tau$, and thus a plant with large $C$ or $R_c$ can effectively slow down dehydration under short-term water stresses and maintain leaf-water potential higher than the threshold which causes stomata to close and  so that photosynthesis and other physiological functioning can proceed. This maintenance of water status is observed in experimental studies of grass leaves. In a leaf that dehydrates slowly (possibly due to larger capacitance), stomatal conductance would change less with atmospheric humidity and VPD by a smaller magnitude of slope compared to a leaf showing faster dehydration dynamics. The two types of water-storage cells in grass leaves, bulliform and bundle sheath, are different in their specific capacitance values $C$, which are related to cell wall rigidity or elasticity, and also in $R_c$ values. Bundle sheaths are much closer to xylem conduits which are contained in vascular bundles, while bulliform cells are mostly distributed in the epidermis on the upper side of a leaf. The longer water pathways from xylem to bulliform may lead to larger $R_c$ and greater contribution to the delay of water loss. In our simulation, using two sets of capacitors and resistors could be feasible when dealing with the cell types. The baseline potential $\psi_s$ of water storage, which does not play an explicit part in this study, can be obtained by measuring $C$ and water content $W=C(\psi_x-\psi_s)$ where $\psi_x$ is the steady-state mean potential in xylem. These measurements can be achieved through pressure-volume (PV) curves, which measure the interdependence between leaf potential and water content \cite{Jones2013}. The obtained $\psi_s$ is determined by a lower limit of osmotic potential (or higher limit of osmotic pressure) due to the presence of solutes in leaf cell water content. Another factor worth considering when studying water flows outside of xylem, both to stomata for transpiration or to leaf cells for storage, is the actual form of water transport. Recent modeling efforts have specifically investigated the relative importance of symplastic (through cell cytoplasm), apoplastic (inside cell wall but outside cell membrane), and gaseous pathways (especially in transpiration), as well as the exact site of exiting water evaporation (either near vascular bundle or near stomata and epidermis) which may complicate the hydraulic condition inside a leaf \cite{Rockwell2014,Buckley2015,Buckley2017}.

\begin{figure}[hbt!]
\centering
\includegraphics[width=0.5\textwidth]{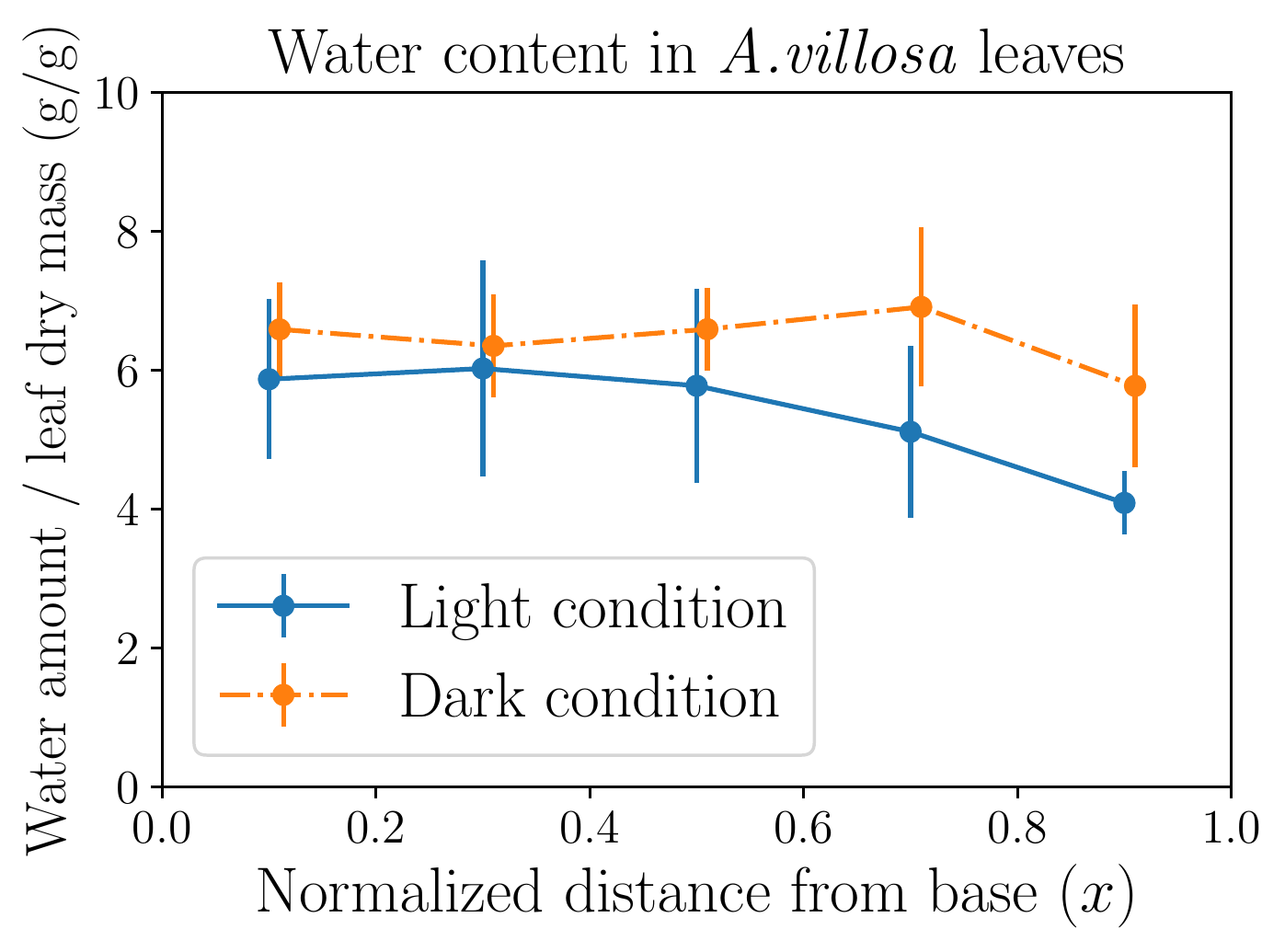}
\caption{The experimental measurements of the mass of water content averaged over leaf dry mass (both in grams), in each of the five leaf portions of equal length. Six leaves of the grass \emph{Anthaenantia villosa} in light conditions and six leaves of the same species in dark conditions are used in the measurements. Error bars represent standard deviations. A more even distribution in leaves in darkness and a tapering trend from base to tip in leaves exposed to light are observed. The two sets of data are plotted with a horizontal shift of 0.01 to show error bars.}
\label{wc}
\end{figure}

Our spatially explicit modeling methods provide a new theoretical approach to the study of fluid dynamics of general flow networks with fluid-storage function. Such networks are not necessarily hydraulic vascular networks found in plant leaves, but could also be found in other water transport systems such as river networks. The applicability of the modeling methods to plant biology is illustrated in the computation and simulation results of grass leaf hydraulics. The use of grass leaves as a model benefits from not only simpler stomata and venation arrangements than dicots, but also the clear presence of water-storage cells (capacitors). The spatially dependent xylem water potential profile, which decreases from base to tip in well-watered steady state (Figure \ref{ss}) when capacitors are static, compares qualitatively well with earlier calculation results of wheat leaves by Altus et al.\ \cite{Altus1985} and most recent experimental measurements (using a novel method) and theoretical predictions of maize leaves by Jain et al.\ \cite{Jain2021}. By tuning the biologically relevant parameters we choose, we can reproduce quantitatively matching results, though these previous works considered local resistances without explicit water storage. The experimental data of the spatial variation of leaf water potential is usually difficult to obtain by using common techniques, while the measurement of local water content distribution is available. Figure \ref{wc} shows the different measurement results of local water amount averaged over leaf dry mass (to account for both leaf area and thickness) in leaves of the grass \emph{Anthaenantia villosa} (\emph{A.villosa}) when exposed to light or in dark conditions. It is observed that the leaves in darkness (where stomata are presumed to be closed) sustain a more evenly distributed water content, while leaves in light conditions (where stomata are presumed to be open) hold more water near the base but gradually decreasing water amount toward the tip, reflecting the declining trend of water potential. Both observations confirm the findings in our modeling, in which closed stomata ($R_a\rightarrow\infty$) in steady state lead to a trivial situation where water flow is stopped and water potential all over xylem is equal to source potential. See supplementary information for experiment details and additional measurement data.

To further improve the ability of our model to accurately predict leaf hydraulic behaviors, we may implement the grass leaf vascular architecture with hierarchy, where major lateral veins and minor intermediate veins are parallel and connected by transverse veins, as shown by Altus et al. In fact, hierarchical structures exist among all plant leaf networks, especially dicotyledonous leaves whose veins are not parallel but instead form branches and loops. It has been experimentally found in dicot leaves that major veins (with high conductivity) are useful for distributing water throughout the leaf blade evenly in a fast manner, and minor veins (with high resistivity) are used to deliver water to leaf cells \cite{Zwieniecki2002}. This observation has also been simulated by a resistor-only model to verify the functions of multiple levels of veins with different resistances in a mesh network \cite{Cochard2004}. Our modeling could help to reveal the function of capacitance in water flow dynamics and balancing of water distribution in similar networked systems. At each level of the hierarchy, the continuous anatomical narrowing of xylem vessels from leaf base to tip can also affect the water potential pattern and can be implemented in our model with a large number of nodes \cite{Lechthaler2020}. Although our modeling results of water stressed conditions (for example in Figure \ref{temp}) does not exactly reproduce the pattern reported by Jain et al., which is likely because of our assumption of constant stomatal conductance as well as evenly distributed capacitance and resistances, we can apply more biologically realistic (and more complicated) anatomical and physiological inputs in future simulation studies based on our current model.

\section*{Acknowledgments}
The authors acknowledge support from NSF-IOS, award 1856587.

\appendix
\section{General analytical solution of the 1D continuous model}
\label{app:general}
In order to solve Equation \eqref{eqn:con}, we separate the xylem water potential into a static part and a time-dependent part, $\psi(x,t)=\psi_u(x)+\psi_t(x,t)$, where the two parts satisfy:
\begin{equation}
\frac{\partial^2\psi_u}{\partial x^2}=\frac{R}{R_a}\psi_u-\frac{R\psi_a}{R_a}
\end{equation}
\begin{equation}
\frac{\partial^3\psi_t}{\partial t\partial x^2}=R(\frac{1}{R_a}+\frac{1}{R_c})\frac{\partial\psi_t}{\partial t}-\frac{1}{CR_c}\frac{\partial^2\psi_t}{\partial x^2}+\frac{R}{CR_c R_a}\psi_t.
\end{equation}
Given the boundary conditions $\psi_u(x=0)$ which is a constant and $I(x=1)=0$, the time-independent $\psi_u$ is solved as:
\begin{equation}
\label{eqn:ss}
\begin{split}
\psi_u(x)=
&\frac{\psi_u(0)-\psi_a}{1+\exp(2\sqrt{R/R_a})}\exp\Big(\sqrt{\frac{R}{R_a}}x\Big)\\
&+\frac{\psi_u(0)-\psi_a}{1+\exp(-2\sqrt{R/R_a})}\exp\Big(-\sqrt{\frac{R}{R_a}}x\Big)+\psi_a.
\end{split}
\end{equation}
The time-dependent $\psi_t$ can be solved through a Fourier transform of the water potential $\tilde{\psi_t}(x,\omega)=1/\sqrt{2\pi}\int\dif t\,e^{-i\omega t}\psi_t(x,t)$ \cite{Asmar2005} which satisfies:
\begin{equation}
i\omega\frac{\partial^2\tilde{\psi_t}}{\partial x^2}=i\omega R\Big(\frac{1}{R_a}+\frac{1}{R_c}\Big)\tilde{\psi_t}-\frac{1}{CR_c}\frac{\partial^2\tilde{\psi_t}}{\partial x^2}+\frac{R}{CR_c R_a}\tilde{\psi_t}
\end{equation}
whose solution is
\begin{equation}
\label{eqn:dyn}
\tilde{\psi_t}(x,\omega)=\alpha(\omega)\exp(\kappa(\omega)x)+\beta(\omega)\exp(-\kappa(\omega)x)
\end{equation}
\begin{equation}
\label{eqn:kappa}
\kappa(\omega)=\sqrt{\frac{i\omega CR(R_c/R_a+1)+R/R_a}{1+i\omega CR_c}}
\end{equation}
where the functions $\alpha(\omega)$ and $\beta(\omega)$ can be determined by given boundary conditions $\psi_t(x=0,t)$ and $I(x=1)=0$. If $\psi(x=0)=\psi_0$ is time-independent, $\psi_t(x=0,t)=0$ and the water status will stay in the steady state $\psi_u$ described by Equation \eqref{eqn:ss}. The steady-state average potential in the xylem is:
\begin{equation}
\label{eqn:barps}
\bar{\psi}=\int_0^1\dif x\,\psi_u(x)=I_0 R_a+\psi_a
\end{equation}
where $I_0=I(x=0)$ is the current entering through the base:
\begin{equation}
\label{eqn:i0}
I_0=-\left.\frac{1}{R}\frac{\partial\psi_u}{\partial x}\right\lvert_{x=0}=\frac{\exp(2\sqrt{R/R_a})-1}{\exp(2\sqrt{R/R_a})+1}\cdot\frac{\psi_{u}(0)-\psi_a}{\sqrt{RR_a}}
\end{equation}
which is equal to the total transpiration current $E$ at steady state.

\section{Analytical solution of the 1D continuous model with an oscillating water potential at base}
\label{app:tide}
In a uniform, continuous, one-dimensional model described by Equation \eqref{eqn:con}, we solved its steady-state water potential distribution $\psi_u(x)$ in Equation \eqref{eqn:ss}. Here we present the solution of an oscillating boundary condition $\psi_t(x=0,t)=A\cos(\omega_0 t+\varphi)$, whose Fourier transform is:
% \begin{equation}
% \begin{split}
% \tilde{\psi_t}(x=0,\omega)
% &=A\sqrt{\frac{\pi}{2}}\{\cos\varphi[\delta(\omega-\omega_0)+\delta(\omega+\omega_0)]+i\sin\varphi[\delta(\omega-\omega_0)-\delta(\omega+\omega_0)]\}\\
% &=A\sqrt{\frac{\pi}{2}}[e^{i\varphi}\delta(\omega-\omega_0)+e^{-i\varphi}\delta(\omega+\omega_0)].
% \end{split}
% \end{equation}
\begin{equation}
\tilde{\psi_t}(x=0,\omega)=A\sqrt{\frac{\pi}{2}}[e^{i\varphi}\delta(\omega-\omega_0)+e^{-i\varphi}\delta(\omega+\omega_0)]
\end{equation}
where $\delta$ is the Dirac delta. From the form of solution $\tilde{\psi_t}(x,\omega)=\alpha(\omega)\exp(\kappa(\omega)x)+\beta(\omega)\exp(-\kappa(\omega)x)$ in Equation \eqref{eqn:dyn}, we have the following boundary conditions: $\alpha(\omega)+\beta(\omega)=A\sqrt{\pi/2}[e^{i\varphi}\delta(\omega-\omega_0)+e^{-i\varphi}\delta(\omega+\omega_0)]$ and $\beta(\omega)\exp(-\kappa(\omega))-\alpha(\omega)\exp(\kappa(\omega))=0$, the latter of which is derived from the fact that $I=-(1/R)\partial\psi/\partial x$ and $I(x=1)=0$. The functions $\alpha$ and $\beta$ are solved to be:
\begin{equation}
\alpha(\omega)=A\sqrt{\frac{\pi}{2}}\frac{e^{i\varphi}\delta(\omega-\omega_0)+e^{-i\varphi}\delta(\omega+\omega_0)}{1+\exp(2\kappa(\omega))}
\end{equation}
\begin{equation}
\beta(\omega)=A\sqrt{\frac{\pi}{2}}\frac{e^{i\varphi}\delta(\omega-\omega_0)+e^{-i\varphi}\delta(\omega+\omega_0)}{1+\exp(-2\kappa(\omega))}.
\end{equation}
We substitute the expressions into $\tilde{\psi_t}(x,\omega)$, which can be inversely transformed into the real time ($\psi_t=1/\sqrt{2\pi}\int\dif\omega\,e^{i\omega t}\tilde{\psi_t}$):
% \begin{equation}
% \label{eqn:osc}
% \begin{split}
% \psi_t(x,t)
% &=\frac{1}{\sqrt{2\pi}}\int\dif\omega\,e^{i\omega t}\tilde{\psi_t}(x,\omega)\\
% &=\frac{A}{2}\int\dif\omega\,e^{i\omega t}\Big[\frac{\exp(\kappa(\omega)x)}{1+\exp(2\kappa(\omega))}+\frac{\exp(-\kappa(\omega)x)}{1+\exp(-2\kappa(\omega))}\Big][e^{i\varphi}\delta(\omega-\omega_0)+e^{-i\varphi}\delta(\omega+\omega_0)]\\
% &=\frac{A}{2}\Big\{e^{i(\omega_0 t+\varphi)}\left[\frac{\exp(\kappa(\omega_0)x)}{1+\exp(2\kappa(\omega_0))}+\frac{\exp(-\kappa(\omega_0)x)}{1+\exp(-2\kappa(\omega_0))}\right]\\
% &\quad+e^{-i(\omega_0 t+\varphi)}\Big[\frac{\exp(\kappa(-\omega_0)x)}{1+\exp(2\kappa(-\omega_0))}+\frac{\exp(-\kappa(-\omega_0)x)}{1+\exp(-2\kappa(-\omega_0))}\Big]\Big\}
% \end{split}
% \end{equation}
\begin{widetext}
\begin{equation}
\label{eqn:osc}
%\begin{split}
\psi_t(x,t)=\frac{A}{2}\Big\{e^{i(\omega_0 t+\varphi)}\left[\frac{\exp(\kappa(\omega_0)x)}{1+\exp(2\kappa(\omega_0))}+\frac{\exp(-\kappa(\omega_0)x)}{1+\exp(-2\kappa(\omega_0))}\right]+e^{-i(\omega_0 t+\varphi)}\Big[\frac{\exp(\kappa(-\omega_0)x)}{1+\exp(2\kappa(-\omega_0))}+\frac{\exp(-\kappa(-\omega_0)x)}{1+\exp(-2\kappa(-\omega_0))}\Big]\Big\}
%\end{split}
\end{equation}
\end{widetext}
where function $\kappa(\omega)=\sqrt{[i\omega CR(R_c/R_a+1)+R/R_a]/(1+i\omega CR_c)}$ as in Equation \eqref{eqn:kappa}. The total water potential distribution is then $\psi(x,t)=\psi_u(x)+\psi_t(x,t)$ (Equations \eqref{eqn:ss}+\eqref{eqn:osc}).

\section{Analytical calculation of a uniform xylem network removed from plant}
\label{app:cut}
We consider a uniform, continuous xylem model which is at the steady state (Equation \eqref{eqn:ss}) when $t<0$, and is removed from base water source when $t=0$. A closed end immediately forms at $x=0$ which is similar to the terminal at $x=1$. In Figure \ref{uniform} at the first node $i=1$ of the network, $I_{0,1}$ becomes zero instantly and we have $I_{1,2}+I_1^{(a)}+I_1^{(c)}=0$, which means $\partial I_{1,2}/\partial t=-\partial I_1^{(a)}/\partial t-\partial I_1^{(c)}/\partial t$ and that:
\begin{equation}
\frac{\partial I_{1,2}}{\partial t}=-\Big(\frac{1}{R_a}+\frac{1}{R_c}\Big)\frac{\partial\psi_1}{\partial t}\Delta x-\frac{1}{CR_c}\Big(I_{1,2}+\frac{\psi_1-\psi_a}{R_a}\Delta x\Big).
\end{equation}
% \begin{equation}
% \begin{split}
% \frac{\partial I_{1,2}}{\partial t}
% &=-\frac{\partial I_1^{(a)}}{\partial t}-\frac{\partial I_1^{(c)}}{\partial t}=-\Big(\frac{1}{R^{(a)}}+\frac{1}{R^{(c)}}\Big)\frac{\partial\psi_1}{\partial t}+\frac{I_1^{(c)}}{C^{(o)}R^{(c)}}\\
% &=-\Big(\frac{1}{R_a}+\frac{1}{R_c}\Big)\frac{\partial\psi_1}{\partial t}\Delta x-\frac{1}{CR_c}\Big(I_{1,2}+\frac{\psi_1-\psi_a}{R_a}\Delta x\Big).
% \end{split}
% \end{equation}
At the continuous limit, $\Delta x\rightarrow 0$ and $I_{1,2}\rightarrow I(x=0)$, and the equation becomes $\partial I(0)/\partial t=-I(0)/(CR_c)$. Along with the initial condition $I(x=0,t<0)=I_0$ in Equation \eqref{eqn:i0}, we obtain:
\begin{equation}
\label{eqn:i0t}
I(x=0,t)=I_0-I_0\Big(1-\exp(-\frac{t}{CR_c})\Big)H(t)
\end{equation}
where $H$ is the Heaviside step function. This equation shows that the current at $x=0$ does not become zero instantly because of the existence of capacitance. The time-dependent part of $I(x=0,t)$ is $I_t(x=0,t)=-I_0[1-\exp(-t/(CR_c))]H(t)$, whose Fourier transform is:
\begin{equation}
\tilde{I_t}(x=0,\omega)=-I_0\sqrt{\frac{\pi}{2}}\Big(\delta(\omega)+\frac{1}{i\pi\omega(i\omega CR_c+1)}\Big).
\end{equation}
From $I=-(1/R)\partial\psi/\partial x$ and $I(x=1)=0$, we can calculate the Fourier transform of the time-dependent part of xylem water potential $\tilde{\psi_t}(x,\omega)=\alpha(\omega)\exp(\kappa(\omega)x)+\beta(\omega)\exp(-\kappa(\omega)x)$ in which:
\begin{equation}
\label{alpha}
\begin{split}
&\alpha(\omega)=\\
&\quad-\sqrt{\frac{\pi}{2}}\frac{I_0 R}{\kappa(\omega)[\exp(2\kappa(\omega))-1]}\Big(\delta(\omega)+\frac{1}{i\pi\omega(i\omega CR_c+1)}\Big)
\end{split}
\end{equation}
\begin{equation}
\label{beta}
\begin{split}
&\beta(\omega)=\\
&\quad-\sqrt{\frac{\pi}{2}}\frac{I_0 R}{\kappa(\omega)[1-\exp(-2\kappa(\omega))]}\Big(\delta(\omega)+\frac{1}{i\pi\omega(i\omega CR_c+1)}\Big).
\end{split}
\end{equation}

The xylem water potential distribution in the real time is not analytically solvable. However, we can instead calculate the time-dependent part of the average xylem potential, whose Fourier transform is:
% \begin{equation}
% \begin{split}
% \tilde{\psi}_{\text{avg}}(x,\omega)
% &=\int_0^1\dif x\,\tilde{\psi_t}(x,\omega)=\frac{\alpha(\omega)}{\kappa(\omega)}(\exp(\kappa(\omega))-1)+\frac{\beta(\omega)}{\kappa(\omega)}(1-\exp(-\kappa(\omega)))\\
% &=-\sqrt{\frac{\pi}{2}}\frac{I_0 R}{\kappa(\omega)^2}\Big(\delta(\omega)+\frac{1}{i\pi\omega(i\omega CR_c+1)}\Big)\\
% &=-\sqrt{\frac{\pi}{2}}\frac{I_0(1+i\omega CR_c)}{i\omega C(R_c/R_a+1)+1/R_a}\left(\delta(\omega)+\frac{1}{i\pi\omega(i\omega CR_c+1)}\right).
% \end{split}
% \end{equation}
\begin{equation}
\begin{split}
&\quad\tilde{\psi}_{\text{avg}}(x,\omega)=\int_0^1\dif x\,\tilde{\psi_t}(x,\omega)\\
&=-\sqrt{\frac{\pi}{2}}\frac{I_0(1+i\omega CR_c)}{i\omega C(R_c/R_a+1)+1/R_a}\left(\delta(\omega)+\frac{1}{i\pi\omega(i\omega CR_c+1)}\right).
\end{split}
\end{equation}
The inverse Fourier transform gives
% \begin{equation}
% \begin{split}
% \frac{1}{\sqrt{2\pi}}\int\dif\omega\,e^{i\omega t}\tilde{\psi}_{\text{avg}}(x,\omega)
% &=-\frac{I_0 R_a}{2}-\int\dif\omega\,e^{i\omega t}\frac{I_0}{2i\pi\omega[i\omega C(R_c/R_a+1)+1/R_a]}\\
% &=-\frac{I_0 R_a}{2}-I_0 R_a\Big[1-\exp\Big(-\frac{t}{C(R_c+R_a)}\Big)\Big]H(t).
% \end{split}
% \end{equation}
\begin{equation}
\begin{split}
&\quad\frac{1}{\sqrt{2\pi}}\int\dif\omega\,e^{i\omega t}\tilde{\psi}_{\text{avg}}(x,\omega)\\
&=-\frac{I_0 R_a}{2}-I_0 R_a\Big[1-\exp\Big(-\frac{t}{C(R_c+R_a)}\Big)\Big]H(t).
\end{split}
\end{equation}
Because of the continuity of $\bar{\psi}$ at $t=0$, which is equal to its steady-state value in Equation \eqref{eqn:barps}, the average xylem potential in real time is:
\begin{equation}
\bar{\psi}=\psi_a+I_0 R_a-I_0 R_a\Big[1-\exp\Big(-\frac{t}{C(R_c+R_a)}\Big)\Big]H(t)
\end{equation}
and the total transpiration current is calculated as:
\begin{equation}
\begin{split}
E
&=\int_0^1\dif x\,\frac{\psi(x,t)-\psi_a}{R_a}=\frac{\bar{\psi}-\psi_a}{R_a}\\
&=I_0-I_0\Big[1-\exp\Big(-\frac{t}{C(R_c+R_a)}\Big)\Big]H(t).
\end{split}
\end{equation}
Both equations result in Expressions \eqref{eqn:pst} and \eqref{eqn:it} when $t>0$.

\section{The equivalence of the network model (Figure \ref{uniform}) and the lumped model (Figure \ref{lump})}
\label{app:lump}
In Subsection \ref{sec:lump} we showed equivalence of the two equations in groups \eqref{comp1}, \eqref{comp3} and \eqref{comp4}. To make the equations in \eqref{comp2} also equivalent, we study how $R_x$ is related to $R=\sum_{i}R_{i-1,i}$. We already defined $\psi_p=\psi_0$ and $\psi_x=\bar{\psi}$, and here we use the steady state of a uniform 1D model that was calculated in Subsection \ref{sec:model} to compare the left hand sides of the equations, which are $\psi_u(x=0)-\psi_u(x=1)$ and $\psi_u(x=0)-\bar{\psi}$ in the 1D model. We estimate their comparison by doing:
\vspace*{-0.29cm}
%\begin{widetext}
\begin{equation}
\label{eqn:comps}
\begin{split}
&\quad\frac{\psi_u(x=0)-\psi_u(x=1)}{\psi_u(x=0)-\bar{\psi}}\\
&=\frac{1+\exp(2\sqrt{R/R_a})-2\exp(\sqrt{R/R_a})}{1+\exp(2\sqrt{R/R_a})-\sqrt{R_a/R}(\exp(2\sqrt{R/R_a})-1)}\approx\frac{3}{2}.
\end{split}
\end{equation}

%\end{widetext}
The typical outside-xylem resistance is larger than the resistance in the xylem, i.e.\ $R_a\gtrsim R$, and we obtain the approximate ratio $3/2$. To compare the right hand sides, which in the uniform model are $R\bar{I}$ where $\bar{I}=(\psi(x=0)-\psi(x=1))/R$ and $R_x I_x$ where $I_x=I_0$, we estimate the ratio:
\begin{equation}
\frac{\bar{I}}{I_x}=\sqrt{\frac{R_a}{R}}\frac{\exp(\sqrt{R/R_a})+\exp(-\sqrt{R/R_a})-2}{\exp(\sqrt{R/R_a})-\exp(-\sqrt{R/R_a})}\approx\frac{1}{2}
\end{equation}
provided that $R_a\gtrsim R$. Combining the two pairs of comparisons, we found that for equations in \eqref{comp2} to be compatible with each other, we could have $R_x=R/3$ which is $1/3$ the total xylem resistance from base to tip.

\bibliography{paper}

\end{document}